\documentclass[]{pnastwo}
\usepackage{graphicx}

\url{}
\copyrightyear{2015}
\issuedate{\today}
\volume{Volume}
\issuenumber{Issue}

\begin{document}

\title{The Bulk and The Tail of Minimal Absent Words in Genome Sequences}

\author{
  Erik Aurell \affil{1}{Department of Computational Biology, KTH Royal Institute of Technology, AlbaNova University Center, SE-10691 Stockholm, Sweden}\affil{2}{Department of Information and Computer Science, Aalto University, FI-02150 Espoo, Finland},
  Nicolas Innocenti\affil{1},\and
  Hai-Jun Zhou\affil{3}{State Key Laboratory of Theoretical Physics,
    Institute of Theoretical Physics, Chinese Academy of Sciences, 
    Beijing 100190, China}
}

\contributor{}
\maketitle

\begin{article}
  
\begin{abstract}
  Minimal absent words (MAW) of a genomic sequence are subsequences that are
  absent themselves but the subwords of which are all present in the sequence.
  The characteristic distribution of genomic MAWs as a function of their length
  has been observed to be qualitatively similar for all living organisms,
   the bulk being rather short, and only relatively few being long.
  It has been an open issue whether the reason behind this phenomenon is
  statistical or reflects a biological mechanism, and what biological
  information is contained in absent words.  
  In this work we demonstrate that the bulk can be described by a probabilistic
  model of sampling words from random sequences, while the tail of long MAWs is of biological origin. 
  We introduce the
  novel concept of a core of a minimal absent word, which are sequences present 
  in the genome and closest to a given MAW. We show that in bacteria and yeast
  the cores of the longest MAWs, which exist in two or more copies, 
  are located in highly conserved regions
   the most prominent example being ribosomal
  RNAs (rRNAs). We also show that while the distribution of the cores of long
  MAWs is roughly uniform over these genomes on a coarse-grained level, on
  a more detailed level it is strongly enhanced in 3' untranslated regions
  (UTRs) and, to a lesser extent, also in 5' UTRs. This indicates that MAWs
  and associated MAW cores correspond to fine-tuned evolutionary relationships, 
  and suggest that they can be more widely used as markers for genomic
  complexity. 
\end{abstract}

\keywords{Minimal absent word; copy-mutation evolution model; random sequence}

\abbreviations{
	AW, absent word;
	MAW, minimal absent word;
    rRNA, ribosomal RNA;
    UTR, untranslated region
}

\dropcap{G}enomic sequences are texts in languages shaped by evolution. The
simplest statistical properties of these languages are short-range dependencies,
ranging from single-nucleotide frequencies (GC content) to $k$-step Markov
models, both of which are central to gene prediction and many other
bioinformatic tasks~\cite{DurbinEddyKroghMitchison}.
More complex characteristics, such as abundances of $k$-mers, sub-sequences of
length $k$, have applications to classification of genomic
sequences~\cite{SandbergBrandenErnbergCoster2003,HaoQi2004,Chor2009,LaRosa2015},
and \textit{e.g.} to fast computations of species abundancies in metagenomic
data~\cite{AmirZuk-2011,Amir-2013,Koslicki-2013,Chatterjee-2014}. 

The reverse image of words present are absent words (AWs), subsequences which 
actually cannot be found in a text. In genomics the concept was first introduced
around 15 years ago for fragment assembly~\cite{Mignosi2001,Fici2006}
and for species identification~\cite{Fofanov2002},
and later developed for inter- and intra-species comparisons
~\cite{Hampikian:2007uq,Herold2008,10.1371/journal.pone.0016065,10.1371/journal.pone.0029344,Wu2010596}
as well as for phylogeny construction~\cite{Chairungsee2012109}. 
A practical application is to the design of molecular bar codes such as in the
tagRNA-seq protocol recently introduced by us to distinguish primary and
processed transcripts in bacteria~\cite{Innocenti-2015a}. 
Short sequences or tags are ligated to transcript $5'$ ends, and reads from
processed and primary transcripts can be distinguished \textit{in silico} after
sequencing based on the tags. For this to be possible it is crucial that the
tags do not match any subsequence of the genome under study, \textit{i.e.} that
they correspond to absent words. In a further recent study we also showed that
the same method allows to separate true antisense transcripts from sequencing
artifacts giving a high-fidelity high-throughput antisense transcript discovery
protocol~\cite{Innocenti-2015b}.
In these as in other biotechnological applications there is an interest in
finding short absent words, preferably additionally with some tolerance. 

Minimal absent words (MAW) are absent words which cannot be found by
concatenating a substring to another absent word.
All the subsequences of a MAW are present in the text. MAWs in genomic sequences
have been addressed repeatedly~\cite{Herold2008,10.1371/journal.pone.0016065,10.1371/journal.pone.0029344,Wu2010596,Chairungsee2012109}
as these obviously form a basis for the derived set of all absent words.
Furthermore, while the number of absent words grows exponentially with their
length~\cite{19426495}, because new AWs can be built by adding letters
to other AWs, the number of MAWs for genomes shows a drastically
different behavior, as illustrated below in  Fig.~\ref{fig:absentwordsplot}(a)
and previously reported in the literature
\cite{19426495,10.1371/journal.pone.0016065,10.1371/journal.pone.0029344}.  
The behavior can be summarized as there being one or more shortest minimal
absent word of a length which we will denote $l_0$, a maximum of the
distribution at a length we will denote $l_{mode}$, and a very slow decay of the
distribution for large $l$. In human $l_0$ is equal to $11$, as first found
in~\cite{Hampikian:2007uq}, $l_{mode}$ is equal to $18$, there being about
$2.25$ billion MAWs of that length, while the support of the distribution extends to
around $10^6$ (Fig.~\ref{fig:absentwordsplot}(c) and (d)). The total number of
human MAWs is about eight billion. As already found
in~\cite{10.1371/journal.pone.0016065} the very end of the distribution depends
on the genome assembly; for human Genome assembly GRCh38.p2 the three longest
MAWs are $1 475 836$, $831 973$ and $744 232$ nt in length.
\begin{figure*}[t]
  \centering
  \begin{minipage}{\columnwidth}
    \centering
    \includegraphics[width=\columnwidth]{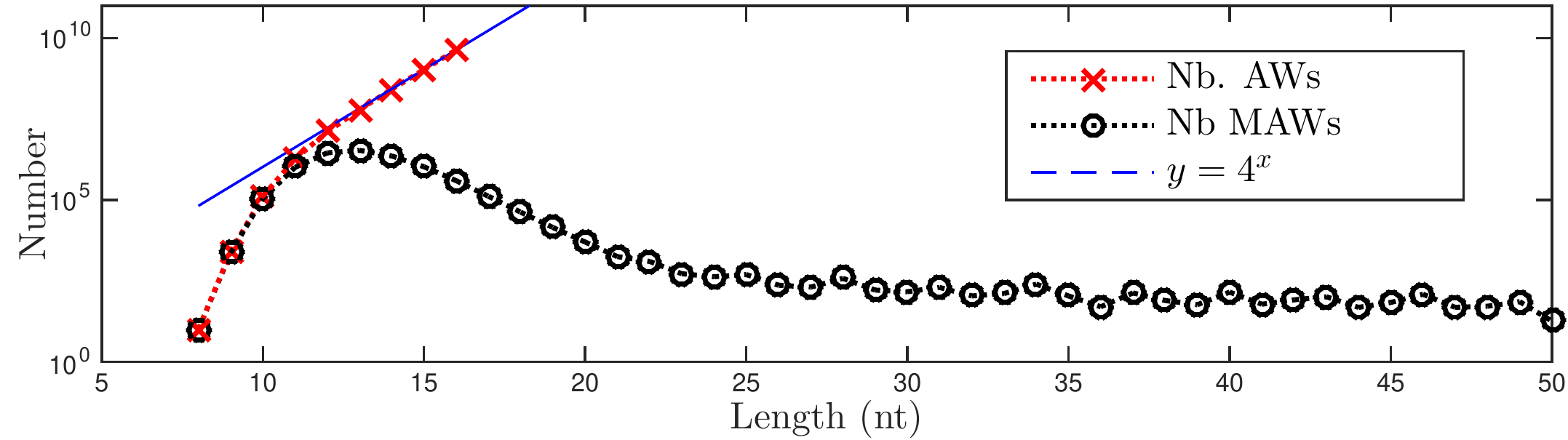} \linebreak
    (a)
  \end{minipage}
  \begin{minipage}{\columnwidth}
    \centering
    \includegraphics[width=\columnwidth]{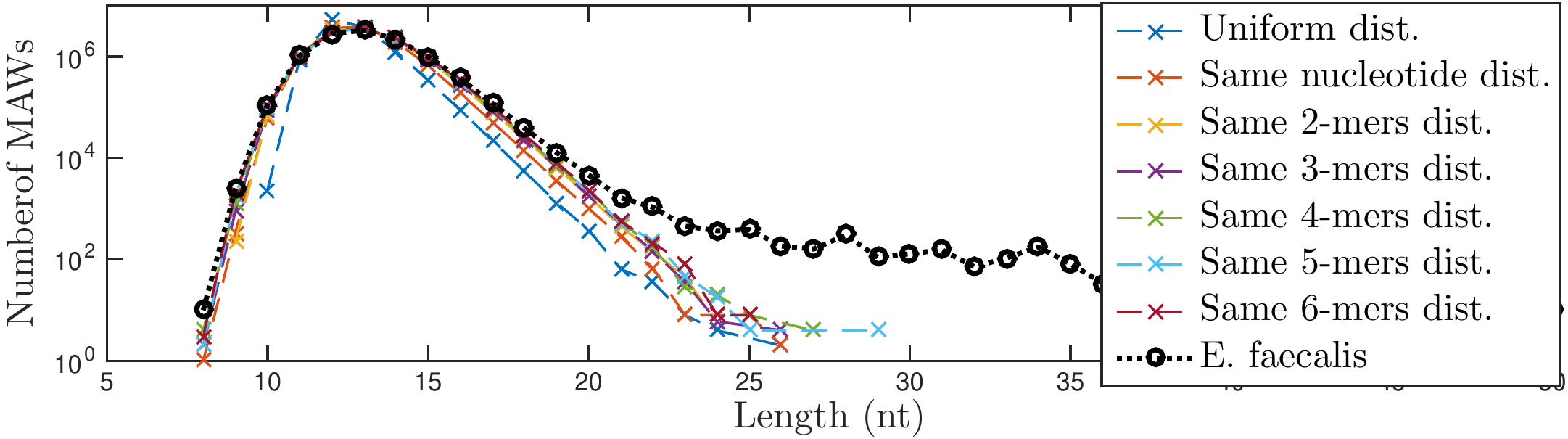}
    \linebreak
    (b)
  \end{minipage}
  \begin{minipage}{\columnwidth}
    \centering
    \includegraphics[width=\columnwidth]{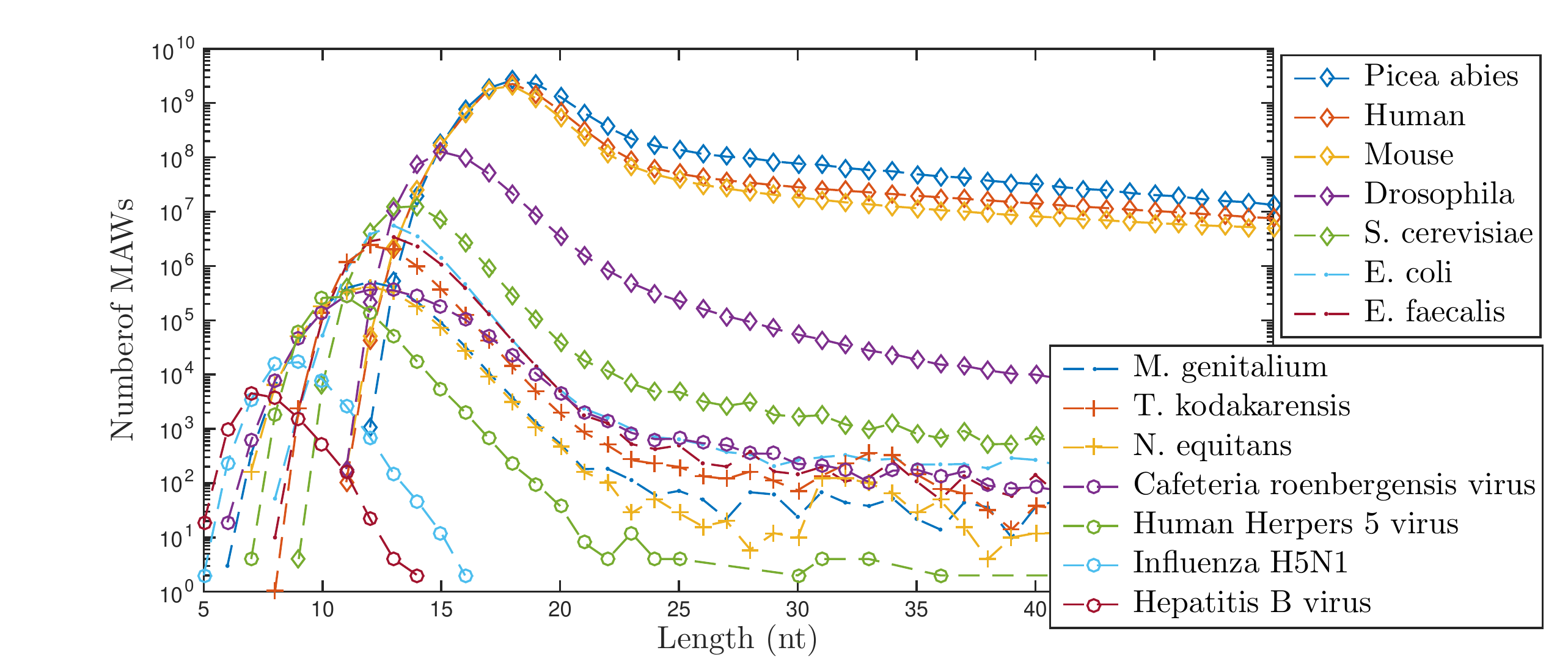}
    \linebreak
    (c)
  \end{minipage}
  \begin{minipage}{\columnwidth}
    \centering
    \includegraphics[width=\columnwidth]{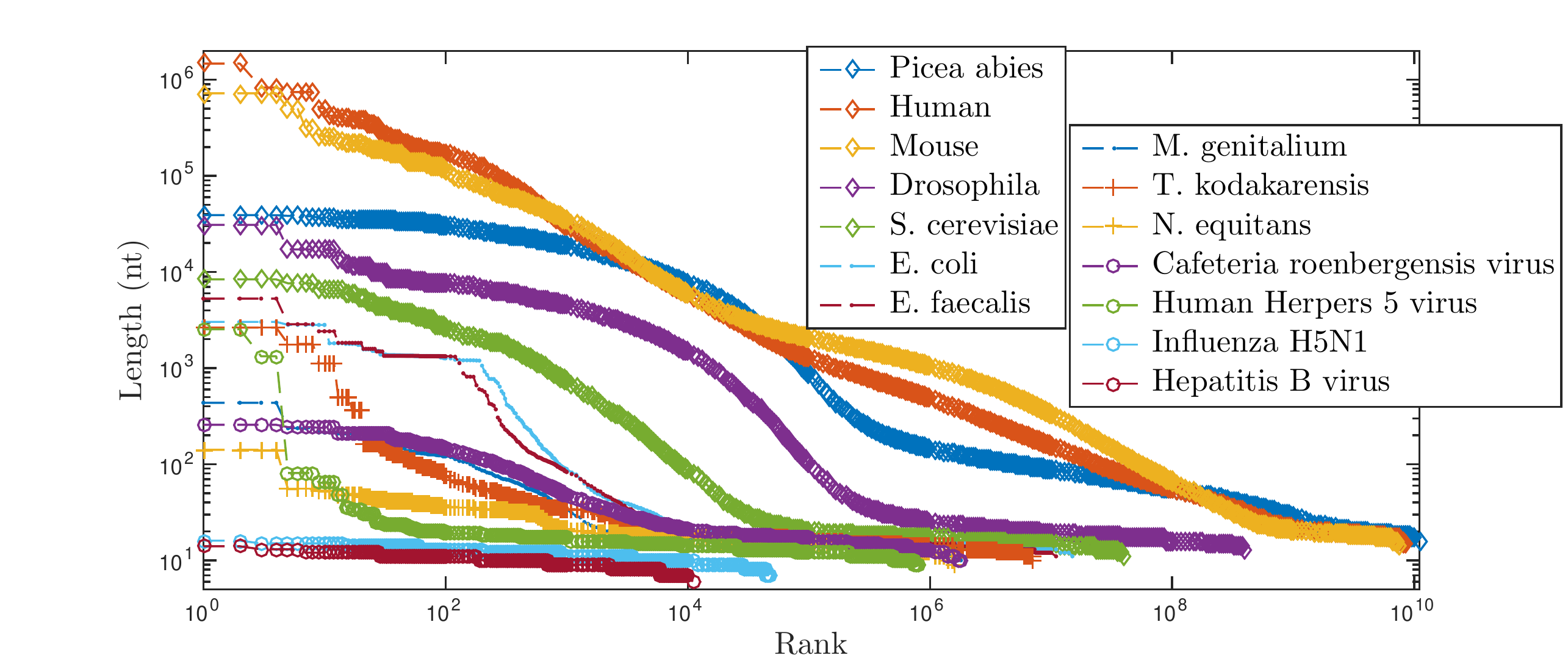}
    \linebreak
    (d)
  \end{minipage}
  \caption{Distributions of the lengths of absent and minimal absent words in genomes and random texts. (a) Number of AWs and MAWs as a function of word length in
    the genome of \textit{E. faecalis} v583.
    The number of AWs grows
    exponentially while the distribution of MAWs shows a maximum and a decay.
    (b) Comparison between the distribution of MAWs in {\it E. faecalis} and the
    ones for a random genome of the same size using different random models.
    (c) Distributions of MAWs for a few common organisms and viruses. 
    (d) Lengths of a MAW as a function of its rank for the distributions shown
    in (c).
    \label{fig:absentwordsplot}
  }
\end{figure*}
Several aspects of this distribution are interesting. First, in a four-letter
alphabet there are $4^k$ possible subsequences of length $k$, but in a text of
length $L$ only $(L-k)$ subsequences of length $k$ actually appear. If the
human genome were a completely random string of letters one would therefore
expect the shortest MAW to be of length $15$. The fact that $l_0$ is
considerably shorter ($11$) is therefore already an indication of a systematic
bias, in~\cite{10.1371/journal.pone.0001022} attributed to the hypermutability
of CpG sites. We will return to this point below. More intriguing is the
observation that the overwhelming majority of the MAWs lie in a smooth
distribution around $l_{mode}$, and then a small minority are found at longer
lengths. We will call the first part of the distribution the \textit{bulk} and
the second the \textit{tail}. We separate the tail from the bulk by a cut-off
$l_{\text{max}}$ which we describe below; the human $l_{\text{max}}$ is $33$, a 
typical number for larger genomes, while the \textit{Escherichia coli}
$l_{\text{max}}$ is $24$. Using this separation there are about $35$ million
human tail MAWs, about $0.447\%$ of the total, while there are $7632$
\textit{E. coli} tail MAWs, about $0.053\%$ of the total. The effect is
qualitatively the same for, as far as we are aware of, all eukaryotic, archeal
and baterial genomes analyzed in the
literature~\cite{19426495,10.1371/journal.pone.0016065}, as well as all tested
by us.  Only a few viruses with short genomes are exceptions to this rule and
show only the bulk, see Fig.~\ref{fig:absentwordsplot}(c).

The questions we want to answer in this work are why the distributions of MAWs 
are described by the bulk and the tail. Can these be understood quantitatively?
Do they carry biological information or are they some kind of sampling effects? 
Can one make further observations? We will show that both the bulk and the tail
can be described probabilistically, but in two very different ways. The bulk of
the MAW distribution arises from sampling words from finite random sequences and
are contained in an interval $[l_{\text{min}},\ l_{\text{max}}]$, where
$l_{\text{max}}$ was introduced above and $l_{\text{min}}$ is a good predictor of
$l_{0}$, the actual length of the shortest AWs. To the best of our
knowledge this has not been shown previously, and although our analysis uses
only elementary considerations, they have to be combined in a somewhat intricate
manner. The distribution of bulk MAWs, which comprise the vast majority of MAWs
in all genomic sequences, can hence be seen as nothing more than a complicated
transformation of simple statistical properties of the sequence. In fact,
excellent results are obtained taking only the single nucleotide composition
into account (Fig~\ref{fig:absentwordsplot}(b)). Nevertheless, the tail MAWs are different, and can be
described by a statistical model of genome growth by a copy--paste--mutate
mechanism similar to the one presented in~\cite{PhysRevLett.90.018101}.
We show that the distributions of the tail MAWs vary, both in the data and in
the model. The human and the mouse MAW tail distributions follow approximately
a power-law, but this seems to be more the exception than the rule; bacteria and
yeast as well as \textit{e.g.} \textit{Picea abies} (Norway spruce)
show a cross-over behavior to a largest MAW length. For bacteria
this largest length ranges from hundreds to thousands; for
\textit{P. abies} it is around 30 000; while for human and mouse the tail MAW
distribution reaches up to one million, without cross-over behavior
(Fig~\ref{fig:absentwordsplot}(d)).

From the definition, any subword obtained by removing letters from the start or
end of a MAW is present in the sequence. In particular, removing
the first and last letters of a MAW leads to a subword that is present at least
twice, which we denote here as a \textit{MAW core}. MAWs made of a repeat of the
same letter are an exception to this rule as they can have the two copies of
their cores overlapping each other, see Appendix A in \textit{Supplemental Information}. MAW
cores can be considered as the causes that create the MAWs and their location on
the genome combined with functional information from the annotation tells
us about their biological significance. Finding all the occurrences on a genome
of a given word (such as a MAW core) is the very common bioinformatic task of
alignment, which can be done quickly and efficiently using one of the many
software packages available. In bacteria and yeast, the cores from the longest
MAWs are predominantly found in regions coding for ribosomal RNAs (rRNAs),
regions present in multiple copies on the genome and under high evolutionary
pressure as their sequence determines their enzymatic properties, required for
protein synthesis and vital to every living cell. At the global scale, it
appears that MAW cores obtained from MAWs in the bulk are distributed roughly
uniformly over the genome while those from the tail cluster in
$3'$ UTRs and, to a lesser extent, also in $5'$ UTRs. These regions are
important for post-transcriptional regulation, and thus likely to be under
evolutionary pressure similarly to rRNAs.

We end this Introduction by noting that from a linguistic perspective a language
can be described by its list of \textit{forbidden} sub-sequences, or the list of
its forbidden words~\cite{HopcroftUllman}
\footnote{In genomics the concept of minimal absent words was first introduced in~\cite{Fici2006},  
  as ``minimal forbidden words''. Since this term has another
  well-established meaning we have here instead used MAW~\cite{19426495}. 
  Related concepts are ``unwords''~\cite{Herold2008} which are the shortest absent words
  (also shortest minimal absent words) and ``nullomers''~\cite{Hampikian:2007uq} which are
  absent words without a requirement on minimality, compare data in Table~1
  in~\cite{Hampikian:2007uq}.}.
Minimal forbidden words relate to
forbidden words as MAWs to absent words, and in a text of infinite length the
lists of MAWs and minimal forbidden words would agree. If there is a finite list
of minimal forbidden words the resulting language lies on the lowest level of
regular languages in the Chomsky
hierarchy\cite{Chomsky1956,ChomskySchutzenberger1963}, and is hence relatively
simple, while a complex set of instructions, such as a genome, is expected to 
correspond to a more complex language, with many layers of meaning. Such aspects
have been exploited in cellular automata theory~\cite{Wolfram1984,Nordahl1989}
and in dynamical systems theory~\cite{Auerbach87}, and are perhaps relevant to
genomics as well. The present investigation is however focused on properties of
texts of finite length, for which minimal forbidden words and MAWs are quite
different. 

\section{A random model for the bulk}
Let us consider a random sequence $\mathcal{S}$ of total length $N$ with
alphabet $\{A, C, G, T\}$. Each position of $\mathcal{S}$ is independently
 assigned the letter $A$, $C$, $G$, or $T$ with corresponding
probabilities $\omega_A$, $\omega_C$, $\omega_G$ and $\omega_T$
($\equiv 1- \omega_A - \omega_C - \omega_G$). A word of length $L$ has the
generic form of ${\bf w} \equiv c_1\, c_2\, \ldots c_{L-1}\, c_L$, 
where $c_i \in \{A, C, G, T\}$ is the letter at the $i$-th position. The total
number of such words is $4^L$. This number exceeds $N$ when $L$ increases to
order $O(\ln N)$, therefore most of the words of length $L\geq O(\ln N)$ will
never appear in $\mathcal{S}$. Then what is the probability $q_{{\bf w}}$ of a
particular word ${\bf w}$ being a MAW of sequence $\mathcal{S}$?

\begin{figure}[b]
  \centering
  \includegraphics[width=.75\columnwidth]{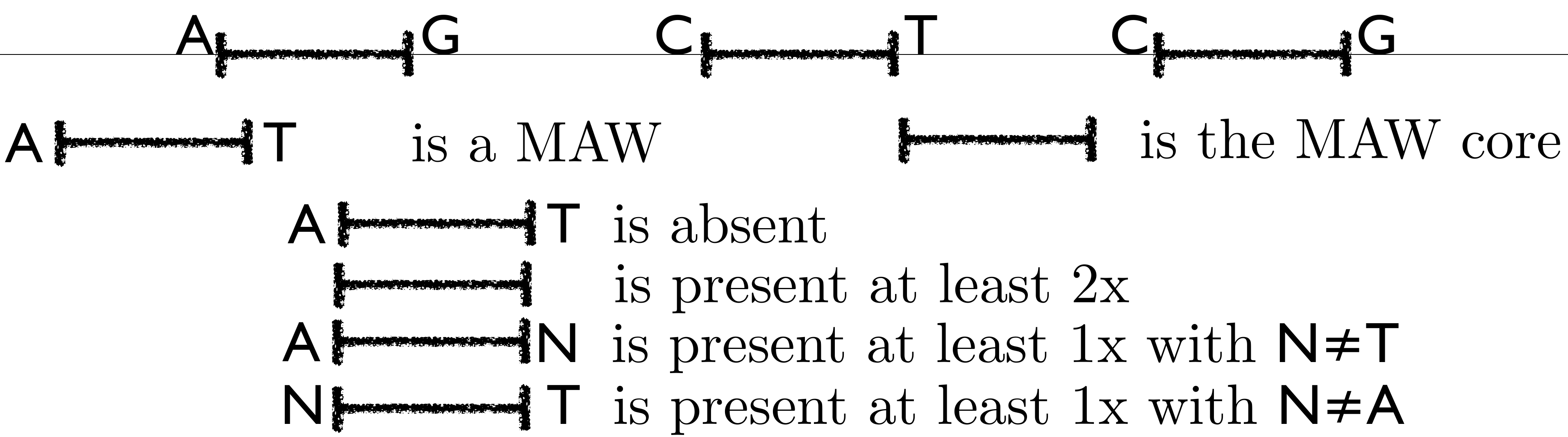}\linebreak
  (a)\linebreak
  \includegraphics[width=\columnwidth]{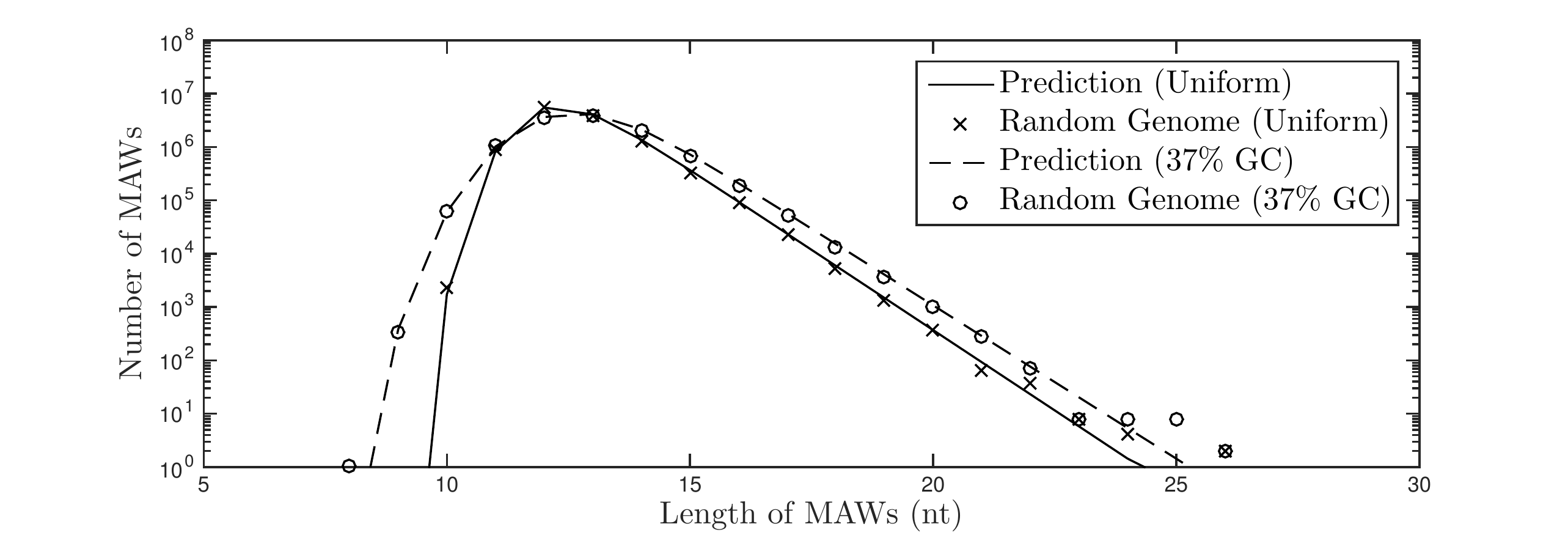} \linebreak
  (b) \linebreak
  \caption{
    \label{fig:ModelvsRandom}
    (a) Illustration of the properties of a minimal absent word and its
    subwords. (b) Comparison between the length distribution predicted by 
    Eq.~\ref{eq:MAWnum} and the number of MAWs calculated for one instance of a
    random genome of $3.3$ Mbp with uniform nucleotide distribution and with
    $37\%$ GC content.
  }
\end{figure}

For ${\bf w}$ to be a MAW, it must not appear in $\mathcal{S}$ but its two
 subwords of length $(L-1)$, ${\bf w}^{(p)} \equiv c_1\, c_2 \ldots c_{L-1}$ and
${\bf w}^{(s)} \equiv c_2 \ldots c_{L-1}\, c_L$,  must appear in $\mathcal{S}$
at least once, as demonstrated schematically in Fig.~\ref{fig:ModelvsRandom}(a).
We define the core of the MAW ${\bf w}$ as the substring
$${\bf w}^{\text{core}} \equiv c_2 \ldots c_{L-1},$$ 
which must appear in $\mathcal{S}$ at least twice, except for the special case
of $c_1=c_2=\ldots = c_{L}$ where the ${\bf w}^{(p)}$ and ${\bf w}^{(p)}$ overlap
(see Appendix A in \textit{Supplemental Information}). The core must immediately follow the
letter $c_1$ at least once and it must also be immediately followed by the
letter $c_L$ at least once. Similarly, if ${\bf w}^{\text{core}}$ immediately
follows the letter $c_1$, it must not be immediately followed by the letter
$c_L$.

We can construct $(N-L+1)$ subsequences of length $L$ from $\mathcal{S}$, say
$\mathcal{S}_1, \mathcal{S}_2, \ldots, \mathcal{S}_{N-L+1}$. Neighboring
subsequences are not fully independent as there is an overlap of length
$(L-m)$ between $\mathcal{S}_n$ and $\mathcal{S}_{n+m}$ with $1 \leq m <L$.
However, for $L \ll N$ two randomly chosen subsequences of length $L$ from the
random sequence $\mathcal{S}$ have a high probability of being completely
uncorrelated. We can thus safely neglect these short-range correlations, and
consequentially the probability of word ${\bf w}$ being a MAW is expressed as
\begin{eqnarray}
  & &\hspace*{-1.0cm} q_{{\bf w}}  =  \bigl[ 1 - \omega({\bf w}) \bigr]^{N-L+1} 
  \nonumber \\
  & & -  \Bigl\{\bigl[ 1 - \omega({\bf w}^{(p)}) \bigr]^{N-L+2} 
  +  \bigl[ 1 - \omega({\bf w}^{(s)}) \bigr]^{N-L+2} 
  \nonumber \\
  & & \quad - \bigl[ 1 - \omega({\bf w}^{(p)}) - \omega({\bf w}^{(s)}) +
    \omega({\bf w}) \bigr]^{N-L+1} \Bigr\} \; ,
  \label{eq:RDprob}
\end{eqnarray}
where $\omega({\bf w}) \equiv \prod_{i=1}^{L} \omega_{c_i}$ (with $c_i$ being the
$i$-th letter of ${\bf w}$) is the probability of a randomly chosen 
subsequence of length $L$ from $\mathcal{S}$ to be identical to the word ${\bf w}$, while
$\omega({\bf w}^{(p)})$ and $\omega({\bf w}^{(s)})$ are, respectively, the 
probabilities of a randomly chosen subsequences of length $(L-1)$ from $\mathcal{S}$
being identical to ${\bf w}^{(p)}$ and ${\bf w}^{(s)}$.

Summing over all the $4^L$ possible words of length $L$, we obtain the expected 
number $\overline{\Omega}(L) \equiv \sum_{{\bf w}} q_{{\bf w}}$ of  MAWs
of length $L$ for a random sequence $\mathcal{S}$ of length $N$:
\begin{eqnarray}
  \overline{\Omega}(L)  &=&
  \sum\limits_{c_1} \sum\limits_{c_L}
  \sum\limits_{n_A, n_C, n_G, n_T} 
  \frac{(L-2)!}{n_A! n_C! n_G! n_T!} \delta_{n_A+n_C+n_G+n_T}^{L-2}
  \nonumber \\
  & & \hspace*{-1.4cm}
  \times \Bigl\{\bigl(1 - \omega_{c_1} \omega_{c_L}
  \omega_A^{n_A} \omega_C^{n_C} \omega_G^{n_G}
  \omega_T^{n_T} \bigr)^{N-L+1} \nonumber \\
  & & \hspace*{-1.0cm}
  -\bigl( 1 - \omega_{c_1} \omega_A^{n_A} \omega_C^{n_C} \omega_G^{n_G} 
  \omega_T^{n_T} \bigr)^{N-L+1} 
  \nonumber 
  \\
  & & \hspace*{-1.0cm}
  -\bigl( 1 - \omega_{c_L} \omega_A^{n_A} \omega_C^{n_C} \omega_G^{n_G} 
  \omega_T^{n_T} \bigr)^{N-L+1}
  \nonumber \\
  & & 
  \hspace*{-1.0cm}
  +\bigl( 1 - (\omega_{c_1} + \omega_{c_L} - \omega_{c_1} \omega_{c_L})
  \omega_A^{n_A} \omega_C^{n_C} \omega_G^{n_G}
  \omega_T^{n_T} \bigr)^{N-L+1} \Bigr\} \; , \nonumber \\
  & &
  \label{eq:MAWnum}
\end{eqnarray}
where the summation is over all the $16$ combinations of the two terminal
letters $c_1, c_L$ and over all the possibilities with which the letters $A$, $C$, $G$, $T$,
may appear in the core a total number of times equal to respectively $n_A$, $n_C$, $n_G$, and $n_T$. 

In the simplest case of maximally random sequences, namely
$\omega_{A} = \omega_C = \omega_G = \omega_T = \frac{1}{4}$, 
Eq.~(\ref{eq:MAWnum}) reduces to
\begin{eqnarray}
  & & \hspace*{-1.0cm} \overline{\Omega}(L) = 4^L (1-4^{-L})^{N-L+1}
  \nonumber \\
  & & \hspace*{-0.7cm} \times
  \Bigl[1- 2 \bigl(1-\frac{3}{4^L-1} \bigr)^{N-L+1} + 
    \bigl(1-\frac{6}{4^L-1}\bigr)^{N-L+1} \Bigr] \; .
  \label{eq:MAWnumMaxRand}
\end{eqnarray}
We have checked by numerical simulations (see Fig.~\ref{fig:ModelvsRandom}b)
that Eq.~\ref{eq:MAWnum} and Eq.~\ref{eq:MAWnumMaxRand} indeed give excellent
predictions of the number MAWs as a function of their length in random sequences.

We define a predicted minimum and a predicted maximum of the support of the bulk
($l_{\text{min}}$ and $l_{\text{max}}$) as the two values of $L$ such that 
$\overline{\Omega}(L)=1$. In the general case, requiring that
$\overline{\Omega}(L) \geq 1$ we obtain $L \geq l_{\text{min}}$, with the
shortest length $l_{\text{min}}$ such that 
\begin{eqnarray}
  & &  \hspace*{-1.0cm}\sum\limits_{n_A, n_C, n_G, n_T} 
  \frac{l_{\text{min}} !}{n_A ! n_C ! n_G ! n_T !}
  \delta_{n_A + n_C + n_G + n_T} ^{l_{\text{min}}} \nonumber \\
  & & \quad \quad \quad \quad \times
  (1-\omega_A^{n_A} \omega_{C}^{n_C} \omega_{G}^{n_G} 
  \omega_{T}^{n_T})^{N-\ell_{\text{min}}} 
  \label{eq:minl-gen}
\end{eqnarray}
is closest to one, while in the other limit we obtain that $L \leq l_{\text{max}}$, with the longest
length $l_{\text{max}}$ being

\begin{equation}
  l_{\text{max}} \approx \frac{2 \ln N}{
    - \ln( \omega_A^2 + \omega_{C}^2 + \omega_{G}^2 + \omega_{T}^2)} \; .
  \label{eq:maxl-gen}
\end{equation}
The bulk distribution is therefore centered around lengths of order $\log N$.
In the case of maximally random sequences, we can obtain the lower limit
analytically and also the first correction to (\ref{eq:maxl-gen}), as
\begin{equation}
  l_{\text{min}} \simeq \frac{\ln N - \ln \ln N}{\ln 4} \; ,
  \label{eq:lmin}
\end{equation}
and
\begin{equation}
  l_{\text{max}} \simeq \frac{2 \ln N + \ln 9}{\ln 4} \; . 
  \label{eq:lmax}
\end{equation}
The above definition of $l_{\text{max}}$ is good enough for our purposes, and
$l_{\text{min}}$ is also a good predictor for $l_0$ (see below). 
A more refined predictor for $l_{\text{min}}$ is discussed in Appendix B in the \textit{Supplemental Information}.  

\section{A random model for the tail}

We now describe a protocol for constructing random genome by a iterative
copy--paste--mutation scheme that qualitatively reproduces the tail behavior
observed for most of real genomes. The model is in principle similar to
\cite{PhysRevLett.90.018101} but differs in the details of the implementation.

The starting point is a string of nucleotides chosen independently at random
with a length $N_0$. At each iteration, we chose two positions $i$ and $j$
uniformly at random on the genome and a length $l$ from a Poisson distribution
with mean $\lambda$. We copy the sequence between $i$ to $(i+l-1)$ and insert it
between positions $j$ and $j+1$, thus increasing the genome size by $l$. We then
randomly alter a fraction $\alpha$ of nucleotides in the genome, choosing the
positions uniformly at random and the new letters from an arbitrary distribution
that can be tuned to adjust the GC content. The process is repeated until the
genome reaches the desired length. 

In this model, $\lambda$ represents the typical size of region involved in a
translocation in the genome and $\alpha$ corresponds to the expected number of
mutations between such events. We observed that the length $N_0$ or the content of the
initial string is unimportant provided that it is much shorter than the final
genome size. The exact value of $\lambda$ given a constant $\alpha/\lambda$ ration 
only affects the tail of the distribution far away from the bulk. We have also checked by simulations that using different
distributions for the choice of $l$ does not
affect the results qualitatively (See Fig.~S1 in \textit{Supplemental Information}). 

\begin{figure}[h]
  \centering
  \includegraphics[width=\columnwidth]{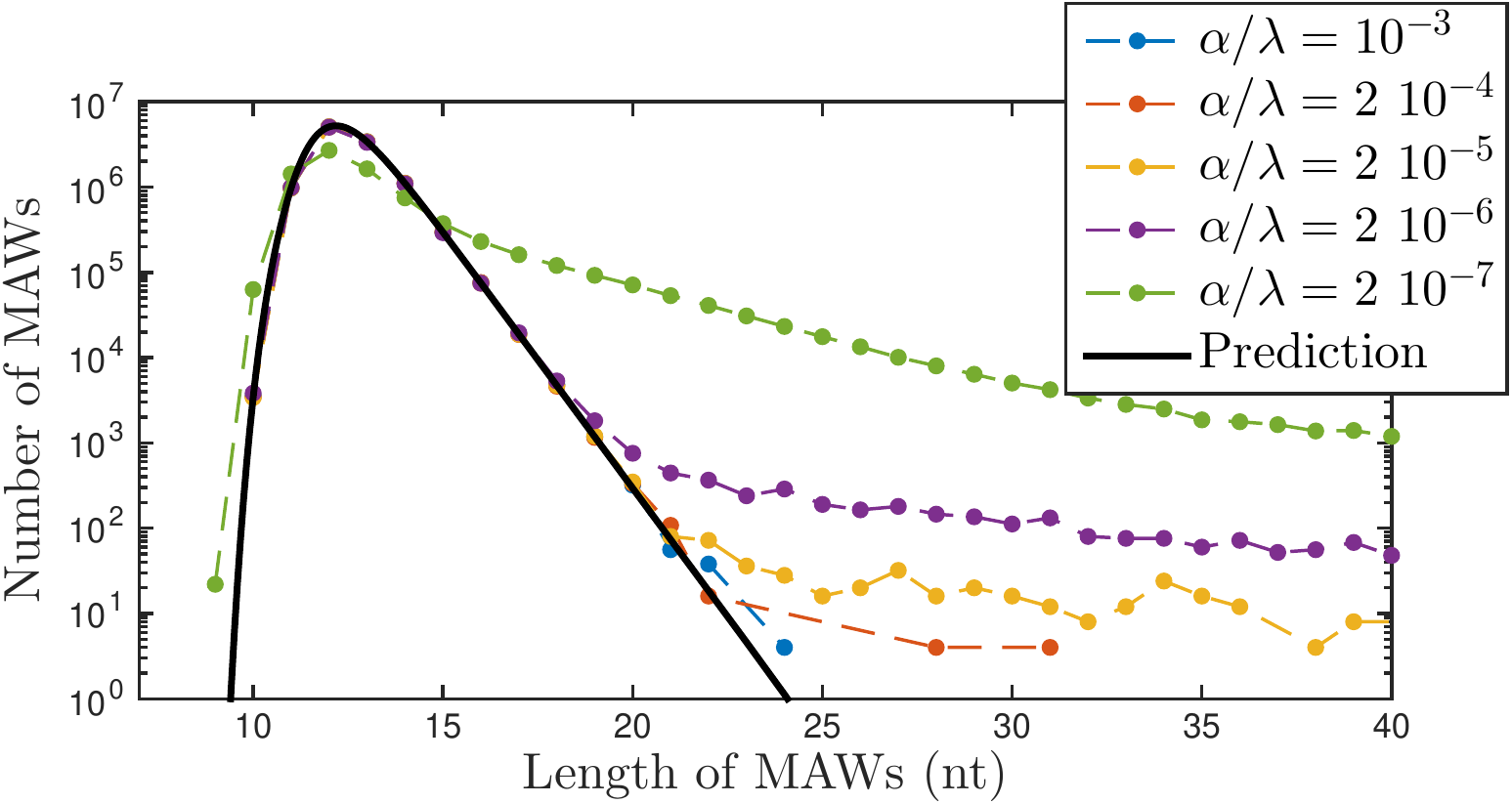}
  \caption{
    Length distributions for MAWs in a few random genomes 3 millions bp in size generated by the
    copy--paste--mutation protocol with different values for the $\alpha/\lambda$ ratio. The curves were obtained using $N_0=5000$ and a constant $\lambda=500$.
    \label{fig:TailModelResult}
  }
\end{figure}

A low ratio $\alpha/\lambda$ generates genomes with tail MAWs while higher
values cause them to only have bulk MAWs as in random texts discussed above
(Fig. \ref{fig:TailModelResult}). This is in agreement with the observations for
viruses in Fig.~\ref{fig:absentwordsplot}: \textit{Hepatitis B} and
\textit{H5N1} are viruses that replicate using an error prone reverse
transcription and the MAW distributions for their genomes lack the tail. In
contrast, \textit{Human Herpers 5} virus and the
\textit{Cafeteria roenbergensis} virus are DNA viruses that use the higher
fidelity DNA replication mechanism and their genomes clearly have tail MAWs.

\section{Estimating the length of the shortest absent words}
\begin{figure}[h]
  \centering
  \includegraphics[width=\columnwidth]{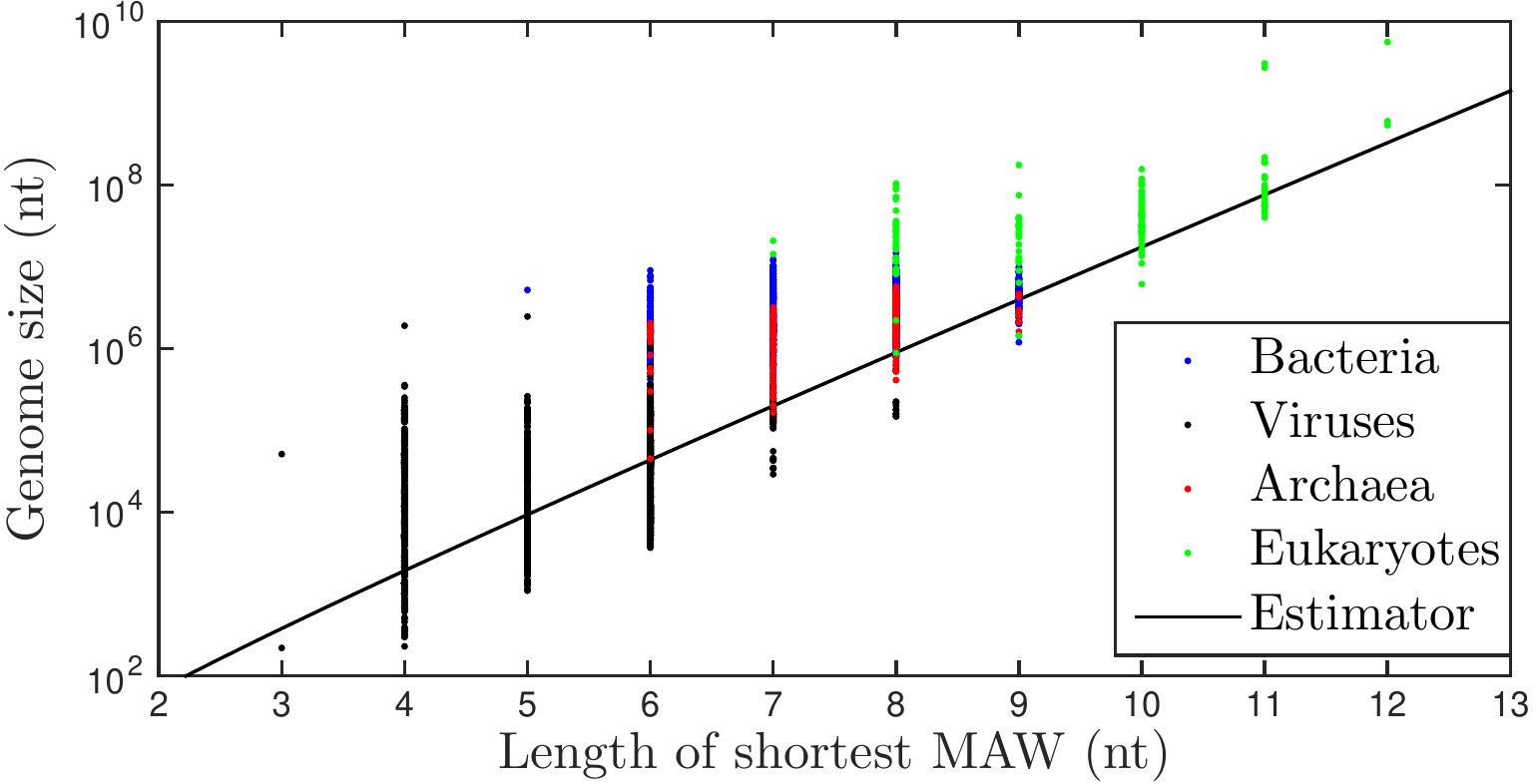}
  \caption{
    Length of the shortest MAW versus the genomes size for all viral, bacterial
    and archaeal genomes available on the NCBI database as well as a few
    arbitrarily chosen eukaryotes, including many short genomes and only a few
    complex organisms, such as Human, mouse and Norway spruce. The black line
    represents the estimator Eq.\ref{eq:lmin}.
    \label{fig:lminAllGenomes}
  }
\end{figure}

Equations (\ref{eq:lmin}) and (\ref{eq:minl-gen}) can be used to estimate the
length of the shortest absent words.  Fig.~\ref{fig:lminAllGenomes} compares the prediction of of the simplest estimate in Eq.~\ref{eq:lmin} to the length of the shortest MAW for a large set of genomes.

The estimator Eq.~\ref{eq:lmin} is expected to be most accurate only for genomes
with neutral GC content. 
The figure reveals that genomes of comparable sizes
typically vary in their $l_0$ by about $4$ nt, and that our estimator captures
very well the upper values in this distribution. 
Using Eq.~\ref{eq:minl-gen} only improves the predictions for genomes 
with much biased GC content ($40\%$ or less) and leads to results in line with the earlier published estimator by Wu et al.~\cite{Wu2010596}, see Appendix C and Table S1 in \textit{Supplemental Information}. 

This analysis shows that, contrary to the conclusion
of~\cite{10.1371/journal.pone.0001022}, there is no need to invoke a biological
mechanism to explain the length of the shortest MAW; it is instead a property of
rare events when sampling from a random distribution. Indeed, the estimator Eq.~\ref{eq:minl-gen} gives the length of the shortest Human MAW
as $12$ and not $15$, only one nucleotide away from the correct answer ($11$).

\section{The origins of tail MAWs}

The Human Herpes virus 5, a double stranded DNA virus with a linear genome of 
$\sim 235.6$ kbp, has four very long MAWs with lengths of $2540$ nt for two of
them, and $1360$ nt for two others, all other MAWs being much shorter ($81$ nt
or less). The cores of these four MAWs come from three regions, two of them
located at the very beginning and very end of the genome, and the third at
position $\sim 195$ kbp made up of the juxtaposition of the reverse complements
of the two others. These regions are annotated as repeated and regulatory. Based
on the NCBI BLAST webservice, these particular sequences are highly conserved
($95\%$ or more) in numerous strains of the virus and do not seem to have
homologues in any other species: the closely related 
\textit{Human Herpes virus 2} shows sequences with no more than $42\%$ 
similarities to these MAW cores.

In \textit{E. coli} and \textit{E. faecalis}, the $10$ longest MAWs with lengths
between $2815$ and $3029$ nt all originate from rRNA regions: a set of genes
present in a few copies made of highly conserved regions with minor variations
between the copies. For yeast, the four longest MAWs ($8376$ nt) originate from
the two copies of rRNA RDN37 on chromosome XII, another four ($7620$ nt) are
caused by $2$ copies of the region containing PAU1 to VTH2 on chromosome X and
PAU4 to VTH2 on chromosome IX and two more ($6531$ nt) originate from the copies
of gene YRF on chromosome VII and XVI (YRF is present in at least $8$ copies on
the yeast genome). 

\begin{figure*}[t]
  \centering
  \includegraphics[width=\textwidth]{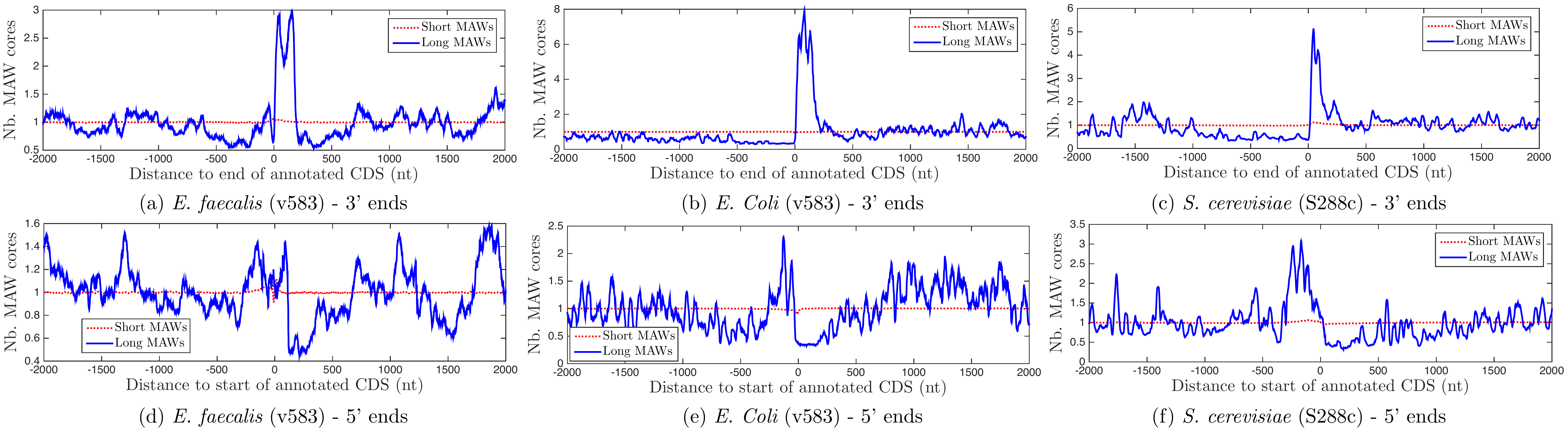}\\
  \caption{
    Average number of MAW cores for MAWs from the bulk and the tail around
    (a)-(c) ends (3' UTRs) and (d)-(f) starts (5' UTRs) of annotated CDS for 
    three living organisms (see also Fig.~S2 in \textit{Supplemental Information}).
    The signals are normalized so that their mean value over the window of
    interest is $1$. The MAW cores concentrate in the UTRs, more strongly on
    the 3' side than on the 5' side.
    \label{fig:distribMAW2}
  }
\end{figure*}

We performed an extensive search for all occurrences of the cores of every MAW
found in the organisms mentioned above and considered the density of MAW cores
along the genome for MAWs in the tail (see Fig. S2 in \textit{Supplemental Information},
). Except for the Human Herpes virus 5 that only has few MAWs which
cluster in the repeated segments discussed above, MAW cores do not appear to be
preferentially located in any specific regions on the genome scale. A more
detailed analysis (Fig. \ref{fig:distribMAW2}) however reveals that, while cores
of MAWs from the bulk appear uniformly distributed, those from the tail cluster
downstream of ends of annotated coding DNA sequences (CDS) (i.e. in the $3'$ 
UTRs and terminator sequences). A similar yet weaker effect can be observed
upstream of the start of annotated CDS (the $5'$ UTR). 
By definition, a MAW core corresponds to a repeated region on the genome
immediately surrounded by nucleotides varying between the copies. Exact repeated
regions lead to only a few MAWs with cores corresponding to that repeat.
Introducing a few random changes in such regions creates more but shorter MAWs,
the cores of which are the sub-strings common to two or more regions. A high
density of MAW cores in a family of regions such as the UTRs thus indicates that
they share a limited set of building blocks, implying a similar set of
evolutionary constraints or a common origin.

\section{The significance of the longest MAWs}

\begin{figure}[b]
  \centering
  \includegraphics[width=\columnwidth]{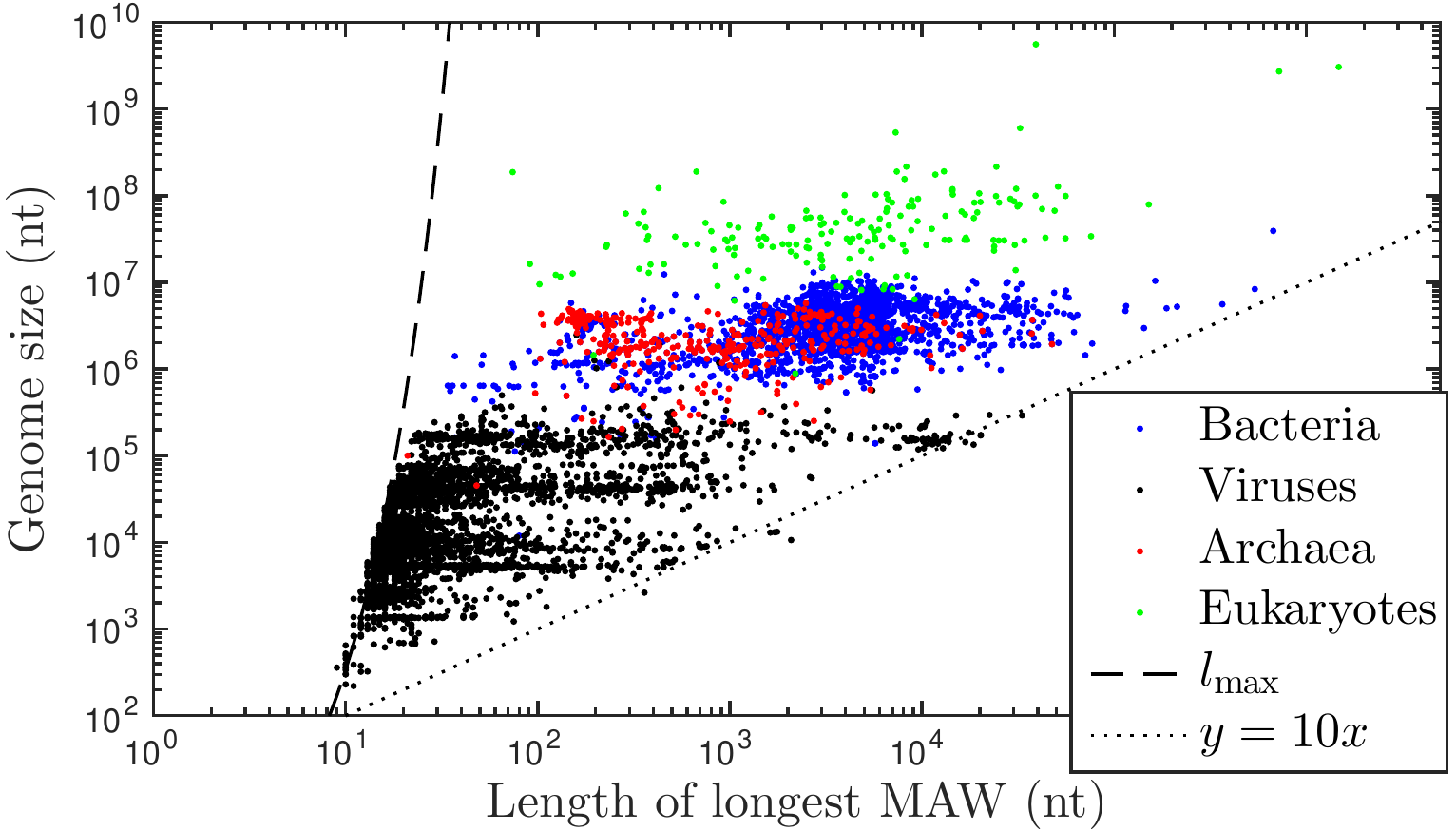}
  \caption{
    Length of the longest absent words as a function of the genome sizes for the same set of
    genomes as in Fig. \ref{fig:lminAllGenomes}. The dashed line represents the
    estimator in Eq.~\ref{eq:lmax} and the dotted line $y=0.1 x$.
    \label{fig:maxlength}
  }
\end{figure}

We now consider the lengths of the longest MAW found in the genomes of numerous 
organisms and viruses. We observe that this length generally lies between 
$l_\text{max}$ and a length of about $10\%$ of the genome size
(Fig.~\ref{fig:maxlength}). 

Viruses are the class showing the largest spread in the length of their longest
MAWs. Many viruses are close to $l_{\text{max}}$, particularly for those with
shorter genomes, confirming our previously mentioned observation that some lack
the tail and are thus closer to random texts. Nevertheless, a few viral genomes
have MAWs longer than $10\%$ of the genome length, \emph{i.e.} proportionally
longer than in any living organism.  The figure suggests that bacteria have on
average slightly longer MAWs than archaea, but overall no clear distinctions
between the four types of genomes can be noted based on the length of MAWs
alone, suggesting that the mechanisms behind evolution of all organisms and
viruses influence the MAWs distribution in the same way.  

A more detailed analysis of the data presented in Fig. \ref{fig:distribMAW2} shows
that organisms have the longest MAWs closest to the lower bound of the tail
($l_{\text{max}}$) or the observed upper bound of $10\%$ of the genome length
share some common traits. 

In bacteria, the $6$ genomes having their longest MAWs closest to
$l_{\text{max}}$ are two strains from the \textit{Buchnera aphidicola} species,
two strains of \textit{Candidatus Carsonella ruddii},
one \textit{Candidatus Phytoplasma solani}
and one \textit{Bacteroides uniformis}.
While the last is a putative bacterial species living in human
feces~\cite{Renouf01062011}, the five other species are intra-cellular symbiotic
or parasitic gammaproteobacteria in plants or
insects~\cite{quaglino2013candidatus,17908294,vanHam21012003}. Among eukaryotes, the same analysis gives
us \textit{Plasmodium gaboni}, an agent responsible for
malaria~\cite{ollomo2009new}, a species of \textit{Cryptosporidium},
another family of intracellular parasites found in drinking water
and \textit{Chromera velia}, a photosynthetic organism from the same
apicomplexa phylum as plasmodium, which is remarkable in this class for its
ability to survive outside a host and is of particular interest for studying the
origin of photosynthesis in
eukaryotes~\cite{keeling2008evolutionary,moore2008photosynthetic}. For Archaea,
we find \textit{Candidatus Parvarchaeum acidiphilum} and
\textit{C. P. acidophilus}, which are two organisms with short genomes ($45.3$
and $100$ kbp) living in low pH drainage water from the Richmond Mine in
Nothern California \cite{baker2010enigmatic}, and an uncultivated hyperthermophilic archaea ``SCGC AAA471-E16'' of the Aigarchaeota phylum \cite{Rinke:2013eu}. 
Additionally, we searched for
MAWs in $2395$ human mithocondrial genomes with lengths between $15436$ and
$16579$ bp and found that the longest MAWs are only $17$ or $18$ nt long, while
$l_\text{max}\simeq 16.5$ for these genomes.

Among bacteria having their longest MAW close to 10\% of the genome length, we
find several strains of \textit{E. coli}, \textit{Francisella tularensis},
\textit{Shewanella baltica}, \textit{Methylobacillus flagellatus},
\textit{Xanthomonas oryzae} and a species of the \textit{Wolbachia} genus. All
of these are facultative or obligatory aerobes and are a lot more widespread
than the bacteria listed in the previous paragraph. At least the four first
species are free-living and commonly cultured in labs. \textit{X. oryzae} is a
pathogen affecting rice residing in the intercellular spaces.
\textit{Wolbachia} species are a very common intracellular parasites or
symbiotes living in arthropods. Among eukaryotes, in addition to human and
mouse, we find \textit{Dictyostelium discoideum}, an organism living in soil
that changes from  uni- to multi-cellular during its life cycle, and 
\textit{Thalassiosira Pseudonana} a unicellular algea commonly found in marine
phytoplankton. Finally, archaea with the longest MAWs are
\textit{Methanococcus voltae}, a mesophilic methanogen,
\textit{Halobacterium salinarum}, a halophilic marine obligate aerobic archeon
also found in food such as salted pork or sausages, and \textit{Halalkalicoccus jeotgali}, another halophilic archeon isolated from
salted sea-food\cite{SGM:/content/journal/ijsem/10.1099/ijs.0.65121-0}.

To summarize, it appears from these results that development in specific
environmental niches that offer rather stable conditions, and particularly the
inside of the cell of another organism, is, albeit with some exceptions,
associated with short MAWs. On the other hand, aerobic life and widespread presence in changing and diverse environments such as soil, sea water or food is
generally associated with long MAWs. 

Intracellular organisms have a well known tendency to reduce the size of their
genomes and increase error rate for DNA replication due to elimination of error
correction mechanisms~\cite{Moran16012009}. This translates into a high value
for $\alpha$ in our random model for the tail and explains why their longest
MAWs are short. In particular, we note that \textit{B. aphidicola} (genome size
of $\sim 650$ kbp), brought out by our analysis, is known to have the highest
mutation rate among all prokaryotes and indeed its longest MAW is as short as
$34$ nt~\cite{Moran16012009,Lynch:2010nr}. As for organisms specialized for a
niche environment, one may hypothesize that proliferation speed is more
important than replication fidelity \cite{Lynch:2010nr}. 

Reasons for a widespread presence in changing environments to increase the
length of the longest MAWs are less clear. Multicellular eukaryotic organisms do
not show particularly high DNA replication fidelity~\cite{Lynch:2010nr}.
The fidelity of \textit{E. coli} is fairly good at about 3.5x the one of human
germline \cite{Lynch:2010nr}, but it is unclear if this is enough to make it
stand out from other bacteria. We speculate that soil, sea water or food, which
are likely to contain many types of microorganisms, may favor species more
likely to undergo horizontal gene transfers, thus increasing the rate of
translocation events and decreasing $\alpha$ in our model, while leaving the
\emph{per generation} mutation rate unchanged. 

\section{Conclusions}

We have proposed a two-parts model explaining the unusual shape of the length
distribution of minimal absent words (MAWs) in genomes. The first part of the
model quantitatively reproduces the bulk of the distribution by considering the
genome as a random text with random and independent letters. The second part is
a stochastic algorithm grounded in basic principles of how genomes evolve,
through translocation events and mutations, that qualitatively reproduces the
behavior in the tail. Our theory provides an estimator for the length of the
shortest MAWs that is remarkably simple and captures well the global trend
observed in large numbers of genomes from all sorts of organisms and viruses.  

Considerations about the longest MAW in a genome reveal sets of organisms sharing
common high-level features such as the type of environment they live in. 
We have shown arguments for believing that organisms and viruses having few tail MAWs
do so because of a low replication fidelity. Why some organisms such as
\textit{E. coli} have long tail MAWs is less clear and replication fidelity
alone does not seem to be a sufficient reason.

Finally, we have introduced the concept of MAW cores, sequences present on the
genome that tell us about what causes the existence of their parent MAW. We
have shown that, while cores from bulk MAWs do not seem biologically relevant,
cores from tail MAWs cluster in regions of the genome with special roles, namely
ribosomal RNAs and untranslated regions surrounding coding regions of genes, a
feature that cannot be explained by a stochastic protocol that ignores the
biological roles of the strings it manipulates.

\begin{materials}

  \section{Data source} 
  Viral and bacterial genomes were downloaded under the form of the "all.fna"
  archives from the "Genomes/Viruses/" and "Genomes/Bacteria" from the NCBI 
  database on 17-18 May 2015 respectively. The Norway spruce's genome
  was downloaded from the ``Spruce genome project''\cite{Nystedt:2013pi} homepage and the yeast genome
  strain 288C\cite{Engel:2014yf} from Saccharomyces Genome Database. Genomes of other eukaryotes
  and archaeas were downloaded from the NCBI database at several different
  dates over the period May-June 2015. The human mithocondrial genomes were
  downloaded from the Human Mitochondrial Genome Database (mtDB)\cite{Ingman01012006} in early
  September 2015. \linebreak
  
   \section{Software \& Computational Ressources} 
  All MAWs were computed using the software provided by Pinho et al. in 
  \cite{19426495}. The software was run taking
  into account the reverse-complementary strand ('-rc' command line switch) and
  requesting MAWs ('-n' command line option) with length up to five million
  nucleotides, i.e. much longer than the expected length of the longest MAWs.
  The search for MAWs was performed on commodity desktop computers for all but
  the Human, mouse, and Norway spruce genomes, for which the computer 
  "Ellen"\footnote{https://www.pdc.kth.se/resources/computers/ellen} from
  the Center for High Performance Computing (PDC) at KTH  was used. \linebreak  
  Localization of occurrences of MAW cores on the genomes for figures \ref{fig:distribMAW2} and S2 was done by aligning
  these subwords to their respective genomes using Bowtie2 
  \cite{Langmead:2012fk}, allowing only strict alignments (command line option 
  '-v 0').  Identical cores from different MAWs are counted independently in
  the coverage. 
\end{materials}

\begin{acknowledgments}
  This research is supported by the Swedish Science Council through
  grant  621-2012-2982  (EA), by  the  Academy  of  Finland  through  
  its  Center  of  Excellence  COIN (EA), and by the Natural Science 
  Foundation of China through grant 11225526 (HJZ). 
  EA thanks the hospitality of KITPC and HJZ thanks the hospitality of 
  KTH.
  The authors thank Profs R\"udiger Urbanke and Nicolas Macris, and 
  Dr. Fran\c{c}oise Wessner for valuable
  discussions, and PDC, the Center for High Performance Computing at KTH, 
  for access to computational resources needed to analyze large genomes. 
\end{acknowledgments}


\end{article}

\end{document}